\documentclass{aa}
\usepackage{natbib}
\usepackage{graphicx}
\usepackage{epsfig}
\newcommand{\teff}{$T_{\rm eff}$}
\newcommand{\logg}{$\log g$}
\newcommand{\feh}{[M/H]}
\newcommand{\dispi}{{\sc dispi}}
\newcommand{\dispis}{{\sc dispi}s}
\newcommand{\ebv}{$E(B-V)$}

\begin{document}
\authorrunning{Willemsen et al.}
\titlerunning{Stellar parameters from \dispis}
\title{Automated Determination of Stellar Parameters from Simulated Dispersed Images for
DIVA}

\author{ P.G. Willemsen\inst{1},
   \ C.A.L Bailer-Jones\inst{2},
    \ T.A. Kaempf\inst{1},
      K.S. de Boer\inst{1}
  }

\offprints{\tt{willemse@astro.uni-bonn.de} }

\institute{ Sternwarte der Universit\"at Bonn, Auf dem H\"ugel 71, D-53121
  Bonn, Germany 
  \and Max-Planck-Institut f\"ur Astronomie, K\"onigstuhl 17, 69117 Heidelberg, Germany
}

\date{ Recieved 4 November 2002 / Accepted 5 February 2003 }

\abstract{We have assessed how well stellar parameters (\teff , \logg\ and \feh) can be
retrieved from low-resolution dispersed images to be obtained by
the DIVA satellite. Although DIVA is primarily an all-sky astrometric mission, it
will also obtain spectrophotometric information for about 13 million stars
(operational limiting magnitude $V \simeq 13.5$ mag). Constructional studies foresee a grating
system yielding a dispersion of $\simeq$ 200nm/mm on the focal plane (first
spectral order). For astrometric reasons there will be no cross dispersion
which results in the overlapping of the first to third  diffraction orders. The 
one-dimensional, \emph{position related} intensity function is called a \dispi\ ({\sc DISP}ersed {\sc I}ntensity).
 We simulated \dispis\ from synthetic spectra taken from \cite{basel97} and \cite{basel98} but for a limited range of metallicites, i.e. our results are for \feh\ in the range $-$0.3 to 1 dex. We show that there is no need to deconvolve these low resolution signals in order to obtain basic stellar
parameters.  Using neural network methods and by including simulated data of
DIVA's UV telescope, we can determine \teff\ to an average accuracy of
about 2 \%  for \dispis\  from stars with 2000 K $\leq$\ \teff\ $\leq$ 20000 K
and visual magnitudes of $V=13$ mag (end of mission data). \logg\ can be
determined for all temperatures with an accuracy better than 0.25 dex for
magnitudes brighter than $V=12$ mag. For low temperature stars with 2000 K
$\leq$\ \teff\ $\leq$ 5000 K and for metallicities in the range $-$0.3 to +1 dex a determination of \feh\ is possible (to better than 0.2 dex) for these magnitudes. For higher temperatures, the metallicity signatures are exceedingly weak at \dispi\ resolutions so that the determination of \feh\ is there not possible.
Additionally we examined the effects of extinction \ebv\ on \dispis\ and found
that  it can be determined to better than 0.07 mag for magnitudes brighter than
$V=14$ mag if the UV information is included.
\keywords{astrometry - stars: fundamental parameters - methods: data analysis - techniques: spectroscopic}} 

\maketitle


\section{Introduction \newline} The DIVA satellite was proposed in 1996 by a
German consortium of astronomical institutes (\citealt{Roeser6}, \citealt{Roeser5})  and is currently foreseen for
launch in 2006.   DIVA will measure the positions, bright\-nesses and proper motions of some 35
million stars. The scientific goal is to study the Milky Way and to
improve the calibration of stellar properties and parameters. This mission
follows up on the  HIPPARCOS satellite which measured parallaxes
for 100\,000 stars. For about 20\,000 of these stars, the accuracy in parallax was
better than 10\%. With the DIVA satellite this number of stars will be increased by at least a factor of 25 (\citealt{Roeser1}).

DIVA will perform an all-sky survey with a limiting visual magnitude of $V
\simeq 15.5$ mag. Note that every observed star will be measured about 120 times in
the course of the mission. The stated magnitude limits refer to the combined
images of all single measurements. The measurements include the precise
determination of positions, trigonometric parallaxes, proper motions, colours
and magnitudes. For about 13 million stars, spectrophotometric data will also be
obtained down to a visual magnitude of $V \simeq 13.5$. An additional
UV telescope will perform photometry in two spectral ranges adjacent to the
Balmer jump. 

\begin{figure}[h!]
\centering
\epsfig{file=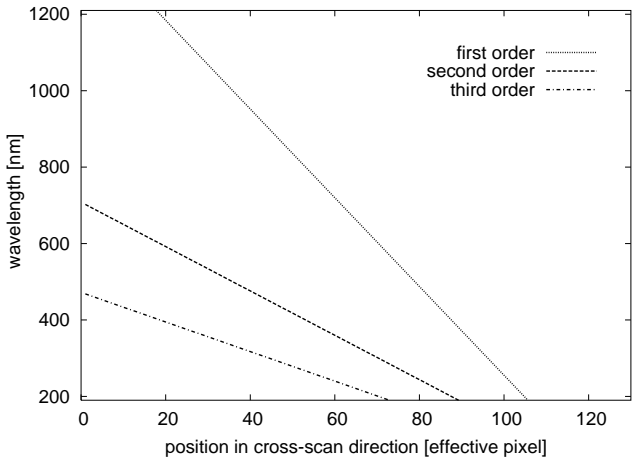,scale = 0.7}
\caption{Wavelength versus position on the spectroscopic CCD (SC) for the first, second and third order. The zeroth
order which is made up of undispersed (white) light would lie at about pixel position 122, but is not shown. 
\label{disp}}
\end{figure}

The DIVA survey represents a large scale and 
deep astrometric and photometric survey of the local part in our Galaxy. The importance of these data to
modern astrophysics will be significant, with applications ranging from stellar
structure and evolution to cosmological aspects. Examples are a precise determination of the luminosity function in
the solar neighbourhood, a better understanding of the structure and formation of our galaxy, the estimation of the amount of dark
matter as well as a better
calibration of the cosmological distance ladder (\citealt{Roeser1}).

After the mission the photometric and spectrophotometric images will be used to
obtain the brightness, the colour and the \dispis\  for the stars. The \dispis\
will allow to derive the astrophysically relevant parameters \teff , \logg ,
\feh\ and \ebv.

Especially the derivation of \logg\ is of importance for objects too distant to
result in an accurate parallax. With these objects in mind, we have carried out
the present study. We will demonstrate that the essential parameters of the
stars can be retrieved with a reasonable level of accuracy from the \dispis\
alone. We will show that astrophysical parameters can be well derived down to the survey limit, perhaps even adequately for stars 1 to 2 magnitudes fainter. There are good scientific arguments to reach fainter in selected fields,  see e.g. \cite{Salim}.

                                                                               
\section{DIVA DISPIS}

\subsection{The concept of a DISPI}
 
 \begin{figure}[h!]
\centering
\epsfig{file=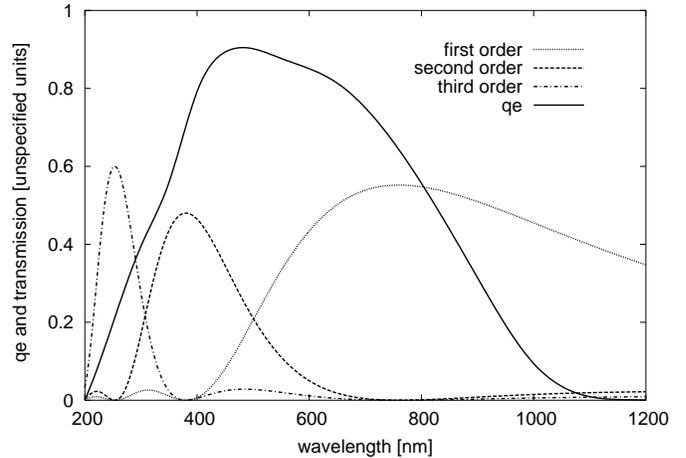,scale = 0.7}
\caption{Transmission curves for the first, second and third grating orders, plus the quantum efficiency.
The order's curves have been scaled down by a factor of 0.6 to better represent data from the real grating.
\label{trans}}
\end{figure}
 
The DIVA satellite is not only unique in its applications and abilities but
also in the way it records spectra.
DIVA will use a grating system yielding a dispersion of
$\simeq$ 200nm/mm on the focal plane with a total efficiency of about 60\%. For
astrometric and other reasons, the resulting (spectral) orders of the grating
are not  separated. Thus ``classical" spectrophotometry will not be obtained.
Instead, the detector will record a pixel related intensity function for each
star, in which all orders (and thus wavelengths) overlap.  Such a
one-dimensional position-coded  intensity distribution is called {\sc dispi}
({\sc DISP}ersed {\sc I}ntensity). Fig.\,\ref{disp} shows the position-coded wavelength of
the grating's orders. One can see that the resolution of the second and third
orders increases by factors of two and three relative to the first order one,
respectively. In the cross-dispersion direction, there are physical pixels while in the direction of dispersion there will be on-chip binning, resulting in ``effective pixels". One such effective pixel corresponds to about 11.6 nm in the first order.

The maximum transmission of the first order is at about 750 nm (see
Fig.\,\ref{trans}), while that of the second and third orders are at about 380 and 250
nm respectively (see also \citealt{Scholz98}, \citealt{Scholz20}). 

\subsection{Dealing with DISPIs}

Given the nature of the \dispis, a classical spectrophotometric analysis -- like line and continuum fitting -- to derive astrophysical parameters
is cumbersome.  We will show that by
training Artificial Neural Networks (ANNs) on simulated \dispis, we can readily
access this information without any further pre-processing of the signal. 
Using \dispis\ from
calibration stars, i.e.\ stars with known physical and apparent properties, we
would initially  build up a standard set of \dispi\ data. The automated classification technique
as developed will then use this library. 
The calibration could be iteratively improved using \dispis\ and their parametrization
results obtained during the mission.
Note that in these simulations, we did not use absolute fluxes as they will be
available from the mission (see below). In this work,
only the shape of a \dispi\ and its line features were  used for tests to
determine basic stellar parameters.

\begin{figure}[h!]
\centering
\epsfig{file=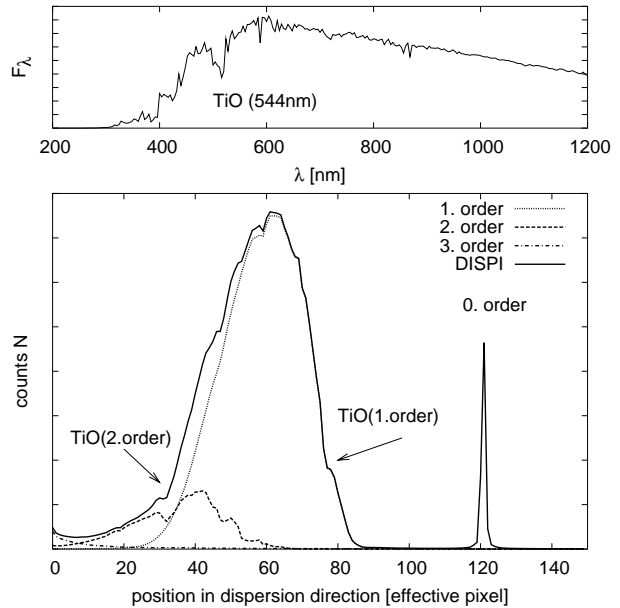,scale = 0.7}
\caption{Input spectrum (top) and simulated DIVA \dispi\ (bottom) for a star with \teff =4500 K, \logg = 4 and solar metallicity (no noise added). Top: the model spectrum sampled in steps of 4 nm, which matches roughly the resolution of the third order of the DIVA dispersed image, shows spectral structure of which the TiO band is marked. Bottom: DIVA \dispi\ showing total counts (in arbitrary units) detected along the effective pixels in the extracted DIVA dispersed information. The TiO feature of the input spectrum (top) can be recognized in the contributions to the first and second orders.
\label{disp4500}} 
\end{figure}

\begin{figure}[h!]
\centering
\epsfig{file=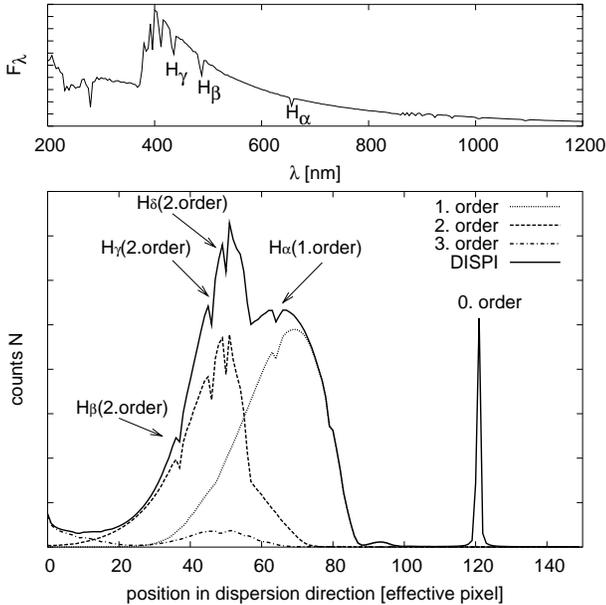,scale = 0.7}
\caption{Input spectrum (top) and simulated DIVA \dispi\ (bottom) for a star with \teff = 9500 K, \logg = 4, solar metallicity (without noise, as in Fig.\,\ref{disp4500}). Here the $H_{\alpha}$, $H_{\beta}$ and $H_{\gamma}$ features are marked in the model spectrum as well as in the resulting \dispi. For such a star, the third order starts to contribute to the signal.
\label{disp9500}} 
\end{figure}

Typical \dispis\ can be seen  in Fig.\,\ref{disp4500} and Fig.\,\ref{disp9500}
 for a cool and a hot star, respectively. One can see that the first, second
 and third orders contribute different amounts to the total light in a \dispi\
 for different temperatures. The first order's transmission maximum is at about
 700 nm while the second and third orders contribute mostly at shorter
 wavelengths. Thus, for the cooler star, the second order contributes less to the \dispi\
 in the case of the bluer, hotter star. The third order's contribution becomes
 negligible for low temperatures. Note that  the ``continuum" of a \dispi\
 is mainly defined by the first and partly by the second order. For a hot star,
 spectral line features are essentially only visible in the second and third
 orders due to their two- and threefold higher resolution. Only for strong molecular
 bands in very cool
 stars are features resolved in the
 first order.

\begin{table}
\centering
\caption[]{The signal-to-noise ratio (S/N) for \emph{single measured} \dispis\ with different temperatures and visual magnitudes as measured from the 10 central pixels around effective pixel position 60 in the direction of dispersion. The innermost 13 TDI-rows of the two-dimensional SC image were summed up to build the \dispi\ (see Sect.\,\ref{input}). The stated temperature is the central value for each sample (sample names are given in parentheses, see Table \ref{set}). The S/N for temperatures in the range of 6000 K $\leq$\ \teff\ $\leq$ 10000 K is almost the same for a given magnitude.
\label{sn}}
\small
\begin{tabular}{cc}
\hline\noalign{\smallskip}
\multicolumn{1}{c}{V[mag]} & \multicolumn{1}{c}{S/N}\\
\noalign{\smallskip}
\hline\noalign{\smallskip}
\multicolumn{2}{c}{\teff = 3000 K ($L_{1}$)}\\ 
\hline\noalign{\smallskip}
8 & 122\\
9& 77\\
10 & 47\\
11 & 28\\
12 & 16\\
\hline
\noalign{\smallskip}
\multicolumn{2}{c}{\teff = 5000 K ($L_{2}$)}\\
\hline\noalign{\smallskip}
8 & 91\\
9 & 56\\
10 & 34\\
11 & 20\\
12 & 11\\
\hline
\noalign{\smallskip}
\multicolumn{2}{c}{\teff = 9000 K ($L_{3}$,$M_{1}$,$M_{2}$)}\\
\hline\noalign{\smallskip}
8 & 82\\
9 & 50\\
10 & 30\\
11 & 17\\
12 & 9\\
\hline
\noalign{\smallskip}
\multicolumn{2}{c}{\teff = 15000 K ($H_{1}$)}\\
\hline\noalign{\smallskip}
8 & 95\\
9 & 59\\
10 & 36\\
11 & 21\\
12 & 12\\
\hline
\noalign{\smallskip}
\multicolumn{2}{c}{\teff = 30000 K ($H_{2}$)}\\
\hline\noalign{\smallskip}
8 & 118\\
9 & 74\\
10 & 46\\
11 & 28\\
12 & 16\\
\hline
\end{tabular}
\end{table} 

The signal-to-noise ratio of a single \dispi\ as measured from the 10 central pixels around effective pixel position 60 is shown for different visual magnitudes and temperatures in Table \ref{sn}.

\section{Artificial Neural Networks}
\label{ANN}
Neural networks have proven useful in a number of scientific
disciplines for interpolating multidimensional data, and thus providing a
nonlinear mapping between an input domain (in this case the \dispis) and an
output domain (the stellar parameters). For an overview of Artifical Neural Networks (ANNs) and 
their application in astronomy for stellar classification see, for example,
\cite{Bailer2002}. The software used in this work is that of \cite{soft}.

A network consists of an input layer, one or two hidden layers and an output
layer. Each layer is made up of several nodes. All the nodes in one layer are
connected to all the nodes in the preceding  and/or following layers. These
connections have adaptable ``weights", so that each node performs a weighted sum
of all its inputs and passes this sum through a nonlinear  transfer function.
That weighted sum is then passed on to the next layer.  Before the network can be
used for parametrisation, it needs to be trained, meaning the weights have to be  set to
their appropriate values to perform the desired mapping. In this process,
\dispis\ together with known stellar parameters as target values are presented
to the network. From these data, the optimum weights are determined by
iteratively adjusting the weights between the layers to minimize an output
error, i.e.\ the discrepancy between the targets and the network outputs. This
is performed by a multidimensional numerical minimization, in this case with
the conjugate gradients method.  When this minimization converges, the weights
are fixed and the network can be used in its ``application" phase: now, only the
\dispi\ input flux vector is presented and the network's outputs produce the
stellar parameters of these \dispis. Since we used only the central 51 effective pixels of the \dispis\ (range 30 to 80, see Fig.\,\ref{disp4500} and \ref{disp9500}), the input layer of
the network was always  made up of the same number of nodes, i.e.\ 51.  We found that the
performance was best when using two hidden layers, each containing 7 nodes.
More nodes did not improve the result significantly but increased the training
time considerably. With four output parameters this network then contains $51 \cdot 7 + 7 \cdot 7 + 7 \cdot 4 = 434$ weights (plus 18 bias weights).

Since we wanted to classify \dispis\ solely based on their shapes, the absolute flux
information was removed by area-normalizing each \dispi, i.e. each flux bin
of a given \dispi\
was divided by the total number of counts in that \dispi.  
Given the non-uniform distribution of the training data over \teff, we classified \dispis\ in terms of
log\,\teff\ instead of \teff.

 Note that, in our tests, we have not included distance information as it
 eventually might be done using DIVA parallaxes, since the present goal was to
 test the retrieval of stellar parameters from \dispis\ only. 

The parametrization errors given below are the average (over some set of \dispis) errors for each parameter, i.e.
\begin{equation} \label{A} A = \frac{1}{N} \cdot \sum_{p=1}^N \left|C(p) -
T(p)\right| \end{equation}  where $p$ denotes the $p^{\rm th}$ \dispi\ and $T$
is the target (or ``true") value for this  parameter. Since the network's
function approximation can depend on the initial settings of the weights, it is
sometimes recommended to use a ``committee" of several networks with identical
topologies but different initializations. The quantity $C(p)$ is the
classification output averaged over a committee of three networks. 


\section{Data simulation}
\label{input}
\subsection{Models of DISPIs}
\begin{figure}[t]              
\centering
\epsfig{file=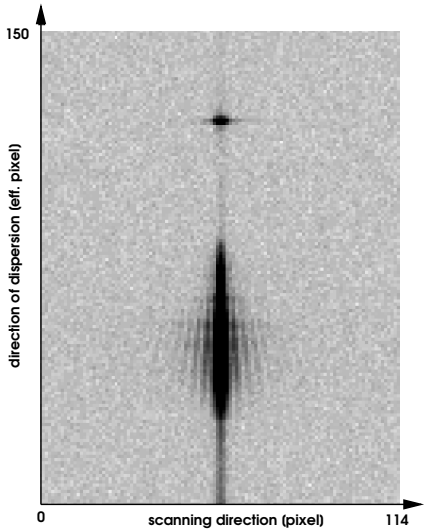,scale = 0.3}
\caption{A dispersed image for an M type star, $V = 10$ mag (with  
noise) as generated by the simulation software from \cite{Scholz98}. Note the contribution of the zeroth order seen as a single intensity 
``blob" in the upper part of the image. The first, second and third spectral order are 
all overlapping (lower intensity ``stretch") due to the grating. The
intensity stripes to the side of the dispersed image are due to diffraction at the
 telescope's aperture.
 \label{fulldisp}}
\end{figure}

The model of the spectrophotometric output from DIVA
used in this work 
was developed by \cite{Scholz98}. This software requires a spectral
energy distribution as input and creates a two-dimensional signal output image
on the detector, containing the dispersed intensity. These images have 114$\times$150 
pixels where the latter number, refering to the dispersion direction, 
is in effective pixels and the former number, refering to the scanning
direction is in physical pixels. Fig.\,\ref{fulldisp} shows such an
image. Ultimately, only a narrow window around the \dispi\ will be read from the focal plane
data stream and
trasmitted to ground, the so-called Spectroscopic (SC) window.

As input spectra we used synthetic spectra from \cite{basel97} and \cite{basel98}. In total there were about 5600 spectra covering a parameter grid
with 68 values for \teff\ between 2000 K and 50000 K (in steps of 200 K for the
low temperature star, and 2500 K for the high temperature stars), 19 possible values for \logg\
ranging from  $-1.02 \leq\ $\logg$\ \leq 5.5$ in steps of approx.\ 0.1 to 0.3 dex
and 13 values for \feh\ with  $-5 \leq\ $\feh$\ \leq 1$ in steps of 0.5 and 0.1
dex. Note that in our tests there were no input data for metallicities in the range from $-$2.5 to $-$0.3 dex.\footnote{In the meantime we started simulations with a set of spectra with a complete range of metallicities.}  

The obvious advantages of using synthetic spectra are the complete wavelength
range from 200 to 1200 nm and the large number of spectra over a large
parameter space. We are currently constructing a library of (previously
published) real stellar spectra. However, since it combines spectra from
many different available catalogues, there is a considerable heterogeneity among
these data. Moreover,  few stars have been observed with the desired wavelength
range from the UV to the IR.  

Interstellar extinction was modelled by using a synthetic extinction curve for $R=3.1$ given as  $\frac{A(\lambda)}{E(B-V)}$ versus $\lambda$.
We used the extinction curve from \cite{Fitzpatrick}, 
simulating 7 different extinction values in steps of 0.15 and 0.2 in the range  $0.0 \leq\ $\ebv$\ \leq 1.0$ mag. Note that the zeroth order was
omitted in this and all other simulations, so that we only worked with dispersed
images made up of the first to third spectral order. Since the data were
area-normalized before passing them through the neural network, the magnitude
information of the zeroth order is lost anyway. For the
simulation of the UV-telescope (see below), the same extinction curve was applied. 
This procedure was done for five different visual magnitudes in the range from
$8 \leq\ $V$\ \leq 12$ mag.

Noise was added to these two-dimensional intensity distributions by passing
them through another software tool developed by Ralf Scholz.
Here, a mean sky backround of $sky = 0.04$e$^-$/(pix s) and a dark current of
$dark = 2$e$^-$/(pix s) were added with additional source and sky Poisson
noise. The CCD's read out noise was 2 e$^-$/eff.pix.

The size of the SC window to be cut from the on-board data stream around the
\dispi\ is crucial as it determines the data rate which is in turn related to
the satellite's overall performance:
A smaller window permits a larger number of SC windows (objects) to be
transmitted. This would, for example, permit a fainter magnitude limit. The optimum window size, i.e.\ the window around a dispersed image with
the highest amount of important and lowest amount of redundant information, was
investigated in earlier studies. Concerning the window size in the cross
dispersion direction it was found that the innermost 7 pixel are sufficient
(\citealt{windivaH})  in terms of highest S/N.   However, due to the satellite's
intrinsic attitude uncertainty it is required that the smallest  acceptable
window size in the scanning direction be 12 pixel (see \citealt{windivaB}). For
our studies we therefore summed up the TDI-rows over the innermost 13 rows (6 pixels in each direction about the
central row). Future work will use a profile fit to
obtain the stellar intensity.

The optimum size in the dispersion direction was evaluated by S/N studies and
the  (spectral) information content. This amount of information was measured by
the ability of Neural Networks to  determine the stellar parameters \teff,
\logg\ and \feh\ for different ranges of \dispis.  It was found
(\citealt{report2}) that these parameters can be adequately retrieved from 
approximately 45 effective pixels around the maximum intensity in the \dispis\
(which is at about effective pixel 60). However, since these earlier studies
included only \dispis\ with \teff $\geq$ 4000 K and since the overall 
intensity distribution moves to smaller effective pixel values for lower
temperatures, we chose the range from 30 to 80 effective pixels in this work.
This should also be appropriate for very red objects like L and M dwarfs with
\teff $\simeq$ 1200--4000 K. 

For further processing, the simulated sky was subtracted from the dispersed
image by evaluating the background level from a single column in scanning
direction next to the dispersed image.

The UV imaging telescope will make use of the same type of CCD's as
the main instrument. The UV magnitudes in the two different passbands next to the
Balmer jump were calculated from the same synthetic spectra as described above,
simply by integrating the flux in the ranges from 310 to 360 nm and 380 to 410
nm. Of course, the true filters will not have exactly square transmission curves,
but this approximation is sufficient for a first analysis of the influence of the UV channel.
The two UV flux values
were fed into the network in three different ways. First, we calculated the
$asinh$ of the flux ratio, i.e. $asinh(UV_{\rm short}/UV_{\rm long}$) (note that the
$asinh$ - function is not undefined for negative values, in contrast to the
log-function. Negative values might occur due to noise for very low temperature
stars with almost no flux in the UV). This ratio is designed to be sensitive to the Balmer jump thus yielding additional information about gravity
and temperature. Second, we summed up the intensity in a \dispi\ in the
range 70 to 80 effective pixel ($\sum_{i=70}^{80}I_i$) and calculated the
ratios ($ UV_{\rm short}/\sum_{i=70}^{80}I_i$) and ($
UV_{\rm long}/\sum_{i=70}^{80}I_i$). Since the first order's contribution in the
selected effective pixel range corresponds to a wavelength range from about 550
to 600 nm (see Fig.\,\ref{disp}), these ratios should be a good measure of
extinction due to the long ``lever" ranging from the UV to the visual/red part
of the spectrum. 

\subsection{Noise in single DISPIs versus end of mission stacked DISPIs}
\begin{figure}[t]
\centering
\epsfig{file=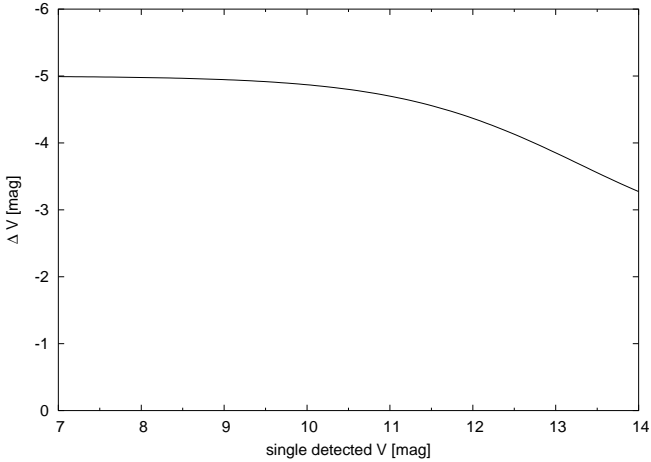,scale = 0.7}
\caption{ One-hundred added frames each of the `single' visual magnitude given on the x-axis yield the same S/N as a star which is $\Delta V$ magnitudes brighter. For example, adding 100 frames of $V=14$ mag stars (two-dimensional intensity distributions) and calculating a \dispi\ from these yields a \dispi\ which has the same S/N as a single \dispi\ of a star with $V=10.8$ mag. Clearly, for fainter stars, the noise is dominated by the read-out noise, while for brighter stars only the Poisson noise of the signal is relevant, thus yielding a full magnitude shift of 5 mag. 
This curve was calculated by making use of the specific DIVA's CCDs noise characteristics.     
\label{final}} 
\end{figure}

The results reported in this paper (Sect.\,\ref{apply}) have been
obtained using \emph{single} \dispis .  However, by the end of the
mission, DIVA will have imaged each star about 120 times. Thus the
final signal-to-noise ratio for any given magnitude will be much
better than from a single measurement. Therefore, the parametrization
performance will also be improved or, equivalently, will be achieved
at a fainter magnitude.  We calculate the final S/N from a sum of 100
two-dimensional intensity distributions. From the ratio of this final
S/N to the single \dispi\ S/N, we can find the equivalent magnitude
difference which gives the magnitude to which our parametrization results
for a {\it single} \dispi\ can be applied to, without having to do a
set of separate simulations on summed \dispis . 
The resulting $\Delta V$ is given in Fig.\,\ref{final} (see further Sect.\,\ref{apply}). We see, for example, that a \dispi\ made up of one-hundred frames each with $V=14$ mag has the same S/N as a single \dispi\ of a star with
magnitude $V=10.8$ mag.  {\it Unless stated otherwise, all results
below will refer to end-of-mission data quality.}

\section{Preparing the ANN input data}
\label{data_sets} 
The ensemble of \dispis\ was divided into several smaller samples with different
temperature ranges. We chose seven different ranges with a broad distinction
between low - (the \emph{L-samples}), mid - (\emph{M-samples}) and  high
temperatures (\emph{H-samples}). The abbreviations as stated in
Table \ref{set} are used throughout this work. The numbers in the last column
show the total number of \dispis\  in the training set in this temperature
interval for the case without and with extinction included (approximately the
same number of \dispis\ in each temperature range was used in the application
set.)

\begin{table}
\caption[]{Abbreviations, temperature ranges and number of \dispis\ in the training sets, with and without extinction (the number in the application sets similar).  
\label{set}}
\begin{flushleft}
\small
\begin{tabular}{ccc}
\hline\noalign{\smallskip}
sample & temperature range & without/with ext.\\
\noalign{\smallskip}
\hline\noalign{\smallskip}
\emph{L$_{1}$} & 2000 K $\leq$ \teff $<$ 4000 K & 330/2300\\
\emph{L$_{2}$} & 4000 K $\leq$ \teff $<$ 6000 K  & 570/3980\\
\emph{L$_{3}$} & 6000 K $\leq$ \teff $<$ 8000 K  & 500/3500\\
\emph{M$_{1}$} & 8000 K $\leq$ \teff $<$ 10000 K  & 400/2800\\
\emph{M$_{2}$} & 10000 K $\leq$ \teff $<$ 12000 K & 180/1200\\
\emph{H$_{1}$} & 12000 K $\leq$ \teff $<$ 20000 K & 390/2700\\
\emph{H$_{2}$} & 20000 K $\leq$ \teff $<$ 50000 K & 450/3100\\
\noalign{\smallskip}
\hline
\end{tabular}
\end{flushleft}
\end{table}

We found that the separation into such small temperature regions yielded  
improved parametrization results. This is understandable as the classification
results for the stellar parameters, especially \logg\ and \feh , depend
upon the presence of spectral features in a \dispi\ which are also closely
related to the temperature of a star.   This effect was also found in
\cite{Weaver2}, using spectra in the near-infrared and classifying
them in terms of MK stellar types and luminosity classes. For our ANN work we chose simple temperature ranges, also aiming at database subsamples of similar size. Though some of the intervals roughly correspond
to the temperatures found in certain MK classes which are characterized by
certain line-ratios, i.e. common physical characteristics, the chosen distinction was
motivated to allow for a reasonable training time for the networks. Another
reason was to see what can be learned from \dispis\ in different temperature
regimes in principle. The mid-temperature samples
(\emph{M}-samples) were defined for the range in which the
Balmer jump  and the H-lines (e.g.\ H$_{\beta}$) change their meaning as
indicators for temperature and surface gravity (see e.g. \citealt{Napiwotzki}). 
Under real conditions one would have to employ a broad classifier to first separate
\dispis\ into smaller (possibly overlapping) temperature ranges. This could be
also based on neural networks, but also on other methods, such as minimum
distance methods.  Each temperature sample was finally divided into two
\emph{disjoint} parts, the training- and the application data. This means that
our classification results (see Sect.\,\ref{apply}) are from \dispis\ in the
gaps of our training grid.

The generalization performance of a network or its ability to classify
previously unseen data is influenced by three factors: The size of
the training set (and how representative it is), the architecture of
the network and the physical complexity of the specific problem, which also
includes the presence of noise. Though there are distribution-free, worst-case
formulae for estimating the minimum size of the training set (based on
the so called VC dimension, see also \citealt{Haykin}), these are often
of little value in practical problems. As a rule of thumb, it is
sometimes stated (\citealt{Duda}) that there should be (W $\cdot$ 10)
different training samples in the training set, W denoting the total
number of free parameters (i.e.\ weights) in the network. In our
network without extinction there were 452 weights. Thus, in some
cases, there were fewer training samples than free
parameters. However, we found good generalization performance (see
Sect.\,\ref{apply} and results therein). This may be due both to (1)
the ``similarity" of the \dispis\ in a specific \teff\ range, giving
rise to a rather smooth (well-behaved) input-output function to be
approximated, and (2) redundancy in the input space. Both give rise to
a smaller number of effective free parameters in the network.

We also tested whether there are significant differences between
determining each parameter separately in different networks and determining all parameters
simultaneously. In the first case each network would have only one ouput node while in the latter case the network had multiple outputs. If the parameters (\teff,
\logg\ etc.)  were independent of each other, one could train a
network for each parameter separately. However, we know that the
stellar parameters influence a stellar energy distribution
simultaneously at least for certain parameter ranges, (e.g.\ hot stars
show metal lines less clearly than cool stars).
Also, for specific spectral features, changes in the chemical composition \feh\
can sometimes mimic gravity effects (see for example \citealt{Gray92}). Varying
extinction can cause changes in the slope of a stellar energy distribution
which are similar to those resulting from a different temperature.

Recently, \cite{Snider2001} determined stellar parameters for low-metallicity
stars from stellar spectra (wavelength range from 3800 to 4500 {{\AA}}). They
reported better classification results when training networks on each parameter
separately. We tested several network topologies with the number of output
nodes ranging from 1 to 3 (in case of extinction from 1 to 4) in different
combinations of the parameters. It was found that 
single output networks did not improve the results. We therefore classified all
parameters simultaneously.                                                                                 
\section{Results } \label{apply}

In this section we report the results from our parameter determination.  In
order to appreciate the results one has to realize that the effects of 
\teff,
\logg\ and \feh\ on a spectrum differ significantly in magnitude. The
strongest signal is that of temperature (the Planck function for black bodies).
The much weaker signal is that of \logg, present in the width of spectral lines
but only weakly in the continuum. Metallicity is a very
weak signal visible in individual spectral lines or perhaps in broader opacity structures (such as G-band or molecular bands). However, in a \dispi\ essentially all line structure is washed out. These general aspects can also be found in \cite{Gray92}. For more see Sect.\,\ref{discussion}. 
\begin{figure*}[t]
\centering
\epsfig{file=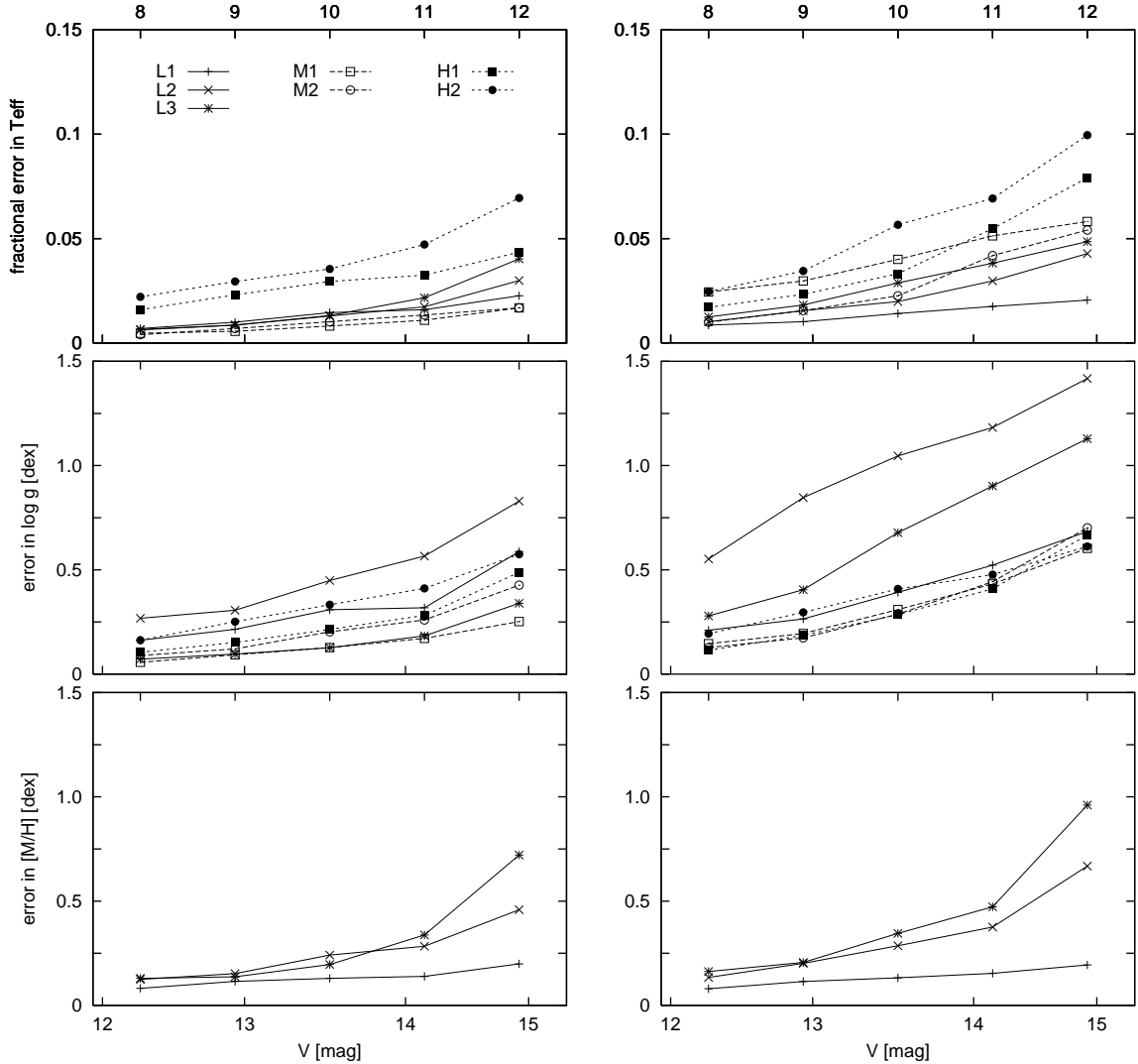,scale = 1.0}
\caption{The average classification uncertainty for \teff , \logg\ and \feh\ from top
to bottom as a function of visual magnitude, V, plotted for the different
temperature sets. No extinction was added for these simulations. The stated
error is that of Eq.\,(\ref{A}).  The lower magnitude scale refers to end of mission data in which 100 \dispis\ of objects with the magnitude given on the upper scale are stacked (see Fig.\,\ref{final} for details). The lettering for the different temperature
ranges refers to the temperature ranges given in Table \ref{set}. The left column shows the
results for networks which were trained on \dispis\ (with an effective pixel
range from 30 to 80) including the UV channel fluxes for the particular star,
while the right column shows results for those simulations without UV data. For \feh\ only the results for the temperature ranges L1 to L3 and
\feh $\geq$ $-$0.3 dex are presented, as for higher temperatures and lower
metallicities spectral information of this parameter is washed out at these
resolutions.
\label{Fno2_ext}} 
\end{figure*}

The errors given are the average errors as in Eq.\,(\ref{A}). 
\teff\ was classified in terms of log\,\teff\ but for better understanding, the resulting errors were transformed to give the \emph{fractional} error in \teff. These fractional errors are stated throughout this paper.

The networks had the same overall topology for all tests, though the number of
inputs naturally was three larger for the case with UV information.
We did not tune the networks to the very best
performance possible. Tests showed that results in individual \teff-ranges
could be improved through adjustments of several free parameters in the
networks (e.g. the number of hidden nodes, number of iterations etc.). In some
cases, we had to adjust some of these parameters, for example in the cases that
the networks did not converge properly. However, such individual tests are very
time-consuming. For the purpose of this paper, we used one topology.

Fig.\,\ref{Fno2_ext} shows the classification results for the stellar
parameters \teff , \logg\ and \feh\ in case of no extinction, while
Fig.\,\ref{G_ext} presents the results for simulations with extinction
included.  For each plot, the left column shows the results when UV data were
included, while the right column refers to the cases without additional UV
data.  The upper magnitude scale shows the visual magnitudes for a single
detection, whereas the lower magnitude scale is relevant for a sum of 100
\dispis, representative of end-of-mission quality data (see Fig.\,\ref{final}).

As a comparison, we tested the performances of random, i.e. untrained, networks.
These are presented in Table \ref{randomtab}. If trained networks give parameters with uncertainites larger than the values listed here, those parameter values are not meaningful. The corresponding error for \ebv\ was about 0.25 mag for all temperature ranges and magnitudes.

\begin{table}
\caption[]{Performances of random (untrained) networks. The stated errors are the average errors as in Eq.\,(\ref{A}) (the error for log\,\teff\ is multiplied by 2.3 to give the fractional error). The errors only depend on the output parameters and are therefore the same for all magnitudes. If trained networks give parameters with uncertainites larger than the values listed here, those parameter values are not meaningful.
\label{randomtab}}
\begin{flushleft}
\small
\begin{tabular}{lccccccc}
\hline\noalign{\smallskip}
Param. & $A_{L1}^{*}$ & $A_{L2}$ & $A_{L3}$ & $A_{M1}$ & $A_{M2}$ & $A_{H1}$ & $A_{H2}$\\
\noalign{\smallskip}
\hline\noalign{\smallskip}
\teff & 0.15 & 0.1 & 0.07 & 0.06 & 0.05 & 0.15 & 0.17\\
\logg & 1.9  & 1.4 & 1.3  & 1.0  & 0.8  & 0.8  & 0.6\\
\feh  & 1.2  & 1.8 & 1.8  &  -   &   -  &   -  &  - \\
\noalign{\smallskip}
\hline
\end{tabular}
$^{*}$ Temperature ranges as in Table \ref{set}
\end{flushleft}
\end{table}

 A variation in \feh\ changes a
 \dispi\ only very subtly for higher temperature stars due to the simple fact that metallicity features are very weak in the energy distribution of hotter stars.


\section{Discussion and conclusions}
\label{discussion}

\begin{figure*}[t]
\centering
\epsfig{file=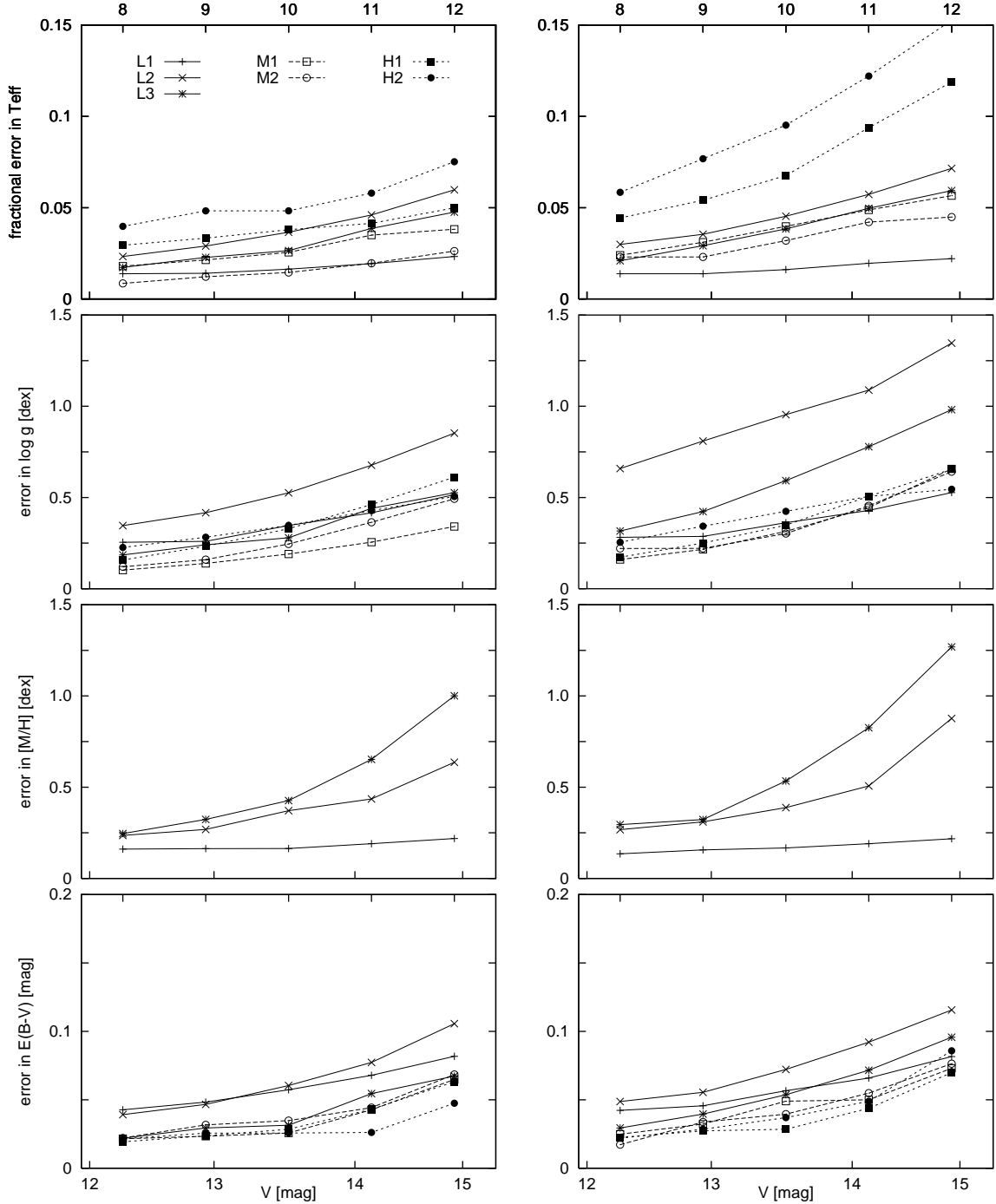,scale = 1.0}
\caption{The same as in Fig.\,\ref{Fno2_ext} but with extinction included. The left column shows the results for networks which were trained with UV data while the right column shows the classification results for tests without UV data.
\label{G_ext}} 
\end{figure*}

As can be seen from Fig.\,\ref{Fno2_ext} and \ref{G_ext} our results show
interesting trends related with the temperature of the stars.  In general,
temperature is the dominating factor for the shape of  and even the details in
a spectral energy distribution.  Overall, \teff\ can be retrieved to very
acceptable accuracy, even without additional UV information: the classification
is better than 10 \% even for very faint stars ($V=14$ mag, no extinction
included: through this section the V magnitude refers to end-of-mission quality data).
Including UV fluxes improves the \teff\ parameter results most noticeably for
hot stars (\teff $>$ 9000 K) and here especially for the fainter ones. For
example, in case of no extinction,  the error for \dispis\  with $V=14$ mag in the \teff\ range 12000--20000 K (H1 sample) drops from 5 \% to about
3 \% when the UV data are included. When extinction is included, the temperature  error drops from 9 \% to about 4 \% for this temperature range at this
magnitude. Though the information in the short
wavelength range is already available in the \dispi , the higher sensitivity of the UV telescope obviously contributes essential information.

Concerning \logg\ we see from the figures that the classification performance
can be improved when additional UV information is included. For example, in case of no extinction and for temperatures in the range 8000 K $\leq$\ \teff\ $<$ 10000 K (M$_{1}$) and visual magniude $V=14$, the error in \logg\ reduces from about 0.4 to 0.15 dex with additional UV information. At this magnitude but for temperatures in the range 6000 K $\leq$\ \teff\ $<$ 8000 K (L$_{3}$), the error reduces considerably from about 0.85 dex to 0.15 dex. These results emphasize the benefit of UV
telescope data in the classification process.  In general, the
\logg\ results are poorer for temperatures in the range
4000 K $\leq$\ \teff\ $<$ 6000 K (L2 sample) when compared to other
ranges.  This is understandable since in this temperature range hardly
any atomic/molecular signatures sensitive to the density (and thus $\log g$) of the
gases in a stellar atmosphere are present.  In contrast, for the very
low temperature ranges the numerous mostly molecular spectral
features provide the information about the density of the atmospheric
gases.  For the higher temperatures the Balmer jump provides still
gravity information.

Metallicity is the most difficult parameter to derive from \dispis. This was to
be expected:   In data with such low spectral resolution  all details of
spectral line information, and thus of metal abundances,  are lost. In Figs.
\ref{Fno2_ext} and \ref{G_ext} only the classification results for
metallicities in the range $-$0.3 dex to 1.0 dex are shown. This is due to a lack of input data in the metallicity range from $-$2.5 dex to $-$0.3 dex (see above). Very low metallicities (\feh\ $\leq$\ $-$2.5 dex) make only a very small imprint on a \dispi\ such that the classification almost fails completely at
these resolutions, except for very bright, cool stars (\teff\ $\leq$\ 4000 K). 
The results for metallicities in the analysed metallicity range are reasonably good even for
low magnitudes (better than about 0.3 dex for $V \leq 14$ mag, no extinction).
When extinction is included, the metallicity performance declines considerably, except for the very cool objects (\teff\ $<$ 4000 K, L1 sample) the
metallicity of which can be determined to better than 0.3 dex for all simulated
magnitudes. 
For \dispis\ in the temperature range 6000 K $\leq$\ \teff\ $<$ 7000 K ($L_{3}$) we see that the error can be reduced from about 0.45 dex to 0.3 dex (no extinction) and from about 0.8 dex to 0.6 dex (with extinction) when the UV information is included. 

Extinction is not easy to be retrieved solely from the shape of a
\dispi , since its overall effect mimics, to some extent, that of
temperature. However, as can be seen from the figures, the
determination of extinction improves when the UV information is
included. This was to be expected since the UV data gives the
strength of the Balmer jump, giving \teff\ information unblurred by
extinction. 

One may ask whether the accuracy of the parameters derived would be
better when working with the single orders of the DIVA dispersed
image. We have tested this for a limited number of parameter
combinations. We generally found that for brighter objects ($V \leq\
13$ mag) the accuracy is improved when using the single orders and then only by a small amount (in the range 30 to 80 effective pixels). This might
be surprising at first but it is understandable since for fainter
objects, using the signal in a single order, deteriorates the S/N, so we find a poorer result from the parameter extraction routine.
Apart from these results, to obtain the separate order's signals would
require a deconconvolution of the \dispi\ which in itself leads to
increased uncertainty in the intensities.  Moreover, a deconvolution
requires knowledge of the nature of the objects so there is no
guarantee that this process will yield unique results.  Since
DIVA will not have separate order images we have not pursued this
aspect further.

Several neural network approaches to stellar parametrization have been
reported in astronomy.  A comparison with those would have to address at least
two aspects,  such as nature of the type of network and characteristics of the data used. 
The networks may indeed be very different in structure such as learning and
regularization technique and especially topology  (compare e.g. \citealt{Weaver1},
\citealt{Snider2001} and this work).  But because in all these cases the data were
of very different nature (e.g. different resolutions and number of input flux
bins),  a general comparison of the results is not really possible.  A few
remarks can be made nevertheless. 

Projects to obtain MK classifications normally use data in the wavelength range
from 3800 to 5200 \AA\ with  high spectral resolution of about 2 to 3 \AA\ (see
e.g. \citealt{Bailer98}). 
\cite{Weaver1} and \cite{Weaver2} classified stars in terms of MK classes in the visual to near-infrared wavelength range 5800-8900 {\AA} with a resolution of 15 {\AA}.  However, even the resolution of these spectra is still much better (by a factor of approx. three) than the ``best" one of \dispis\ which is about 40 {\AA} in the low efficiency third order.  
We would expect such resolutions to give better precision for spectral type or \teff\ as well as for line sensitive parameters (\logg\ and \feh) than with \dispis\ on account of the higher resolution. \cite{Snider2001} recently classified spectra having 1 to 2 {\AA} resolution. They determined \teff ,
\logg\ and \feh\ of low metallicity stars to an accuracy of about 150 K in
\teff\   in the range 4250 K $<$ \teff $<$ 6500 K, 0.30 dex in \logg\ over the
range $1.0 \leq\ $\logg$\ \leq 5.0$ dex and 0.20 dex in \feh\ for $-4 \leq\ $\feh$\ \leq 0.3$ dex. 
From our results, we find for this temperature range ($L_{2}$ and $L_{3}$ sample) a classification precision in \teff\ of better than 5\% for $V \leq 14$ mag (no extinction) without UV information and about 2\% when UV data are included. Only for brighter stars ($V \leq 13$ mag) do we find that \logg\ can be determined from \dispis\ to better than 0.3 dex for temperatures in the range 4000 K $\leq$\ \teff\ $<$ 6000 K but only when UV data are included. Concerning metallicity, our results are comparable (better than 0.2 dex for visual magnitudes $V \leq\ 12$ mag, no extinction, UV data included). Clearly, this is because we have only used metallicities in the range from $-$0.3 dex to +1 dex (see above). 

A comparison with the neural network approach using synthetic data for
a test of possible GAIA photometric systems (\citealt{synspec}) may be of
relevance.  Bailer-Jones also used input data with various moderate
resolutions, some of them similar to those of the spectral orders in
\dispis.  The effects of the quantum efficiency (QE) of
the detectors was not included, so that in his tests the information provided in
the vicinity of (and shortward of) the Balmer jump could be utilized
in full. After shifting our results to the fainter magnitudes reachable 
with GAIAs larger telescope, while considering the other differences 
between Bailer-Jones' and our investigation as well as the differences 
between the DIVA and GAIA optics and data format, 
one must conclude that these ANN analyses work to similar satisfaction. 

Little work has been done so far concerning the automated determination  of
interstellar extinction. \cite{Weaver1} tested the effect of extinction and found
that \ebv\ could be determined from spectra of A type stars with an accuracy of
0.05 mag in the range of \ebv\ of 0 - 1.5 mag. \cite{Gulati97} used IUE
low-dispersion spectra (wavelength range 1153 - 3201 {\AA}, spectral resolution
6 {\AA}) from O and B stars.  Applying reddening to their spectra in the \ebv\ range of 0.05 - 0.95 mag in steps of 0.05 mag, they were able to retrieve
extinction  with high accuracy to about 0.08 mag, clearly because of the 
presence of the 2200 \AA\ bump in the input data. From Fig.\,\ref{G_ext} we see that extinction can be determined from \dispis\ to better than 0.08 mag for visual magnitudes $V \leq\ 14$ mag for all temperatures in case that UV data are included.

Concerning the DIVA satellite, \cite{Elsner99} found interesting results with
Minimum Distance Methods. However, the optical concept of DIVA has changed considerably since then. Thus, also here a comparison of the quality may lead to a skewed judgement  and we therefore refrain from going into detail. 

A final remark deals with the effect of selecting our training sample randomly
from the database. A random selection may accidentally lead to larger regions
of parameter space without training data. Since our object (``application")
sample is the complement of the training set, objects falling in those gaps
clearly are classified worse than objects near trained points. \cite{Malyuto2}
demonstrated such effects when using Minimum Distance Methods. The average
errors in our results are influenced by such effects, but this was not
investigated.

In considering the accuracies obtained with our ANN approach we have to note that real stellar spectra show a much more complex behaviour than synthetic ones. For example, even the more realistic approach of non-LTE models (\citealt{Hauschildt99}) cannot properly describe  the true behaviour of elements in a stellar atmosphere (see e.g. \citealt{Gray92}, Ch.\,13). Moreover, good colour calibrations in accord with observed data are still difficult to obtain, as described e.g. in \cite{Westera2002}.
It is difficult to estimate how to properly weigh such intrinsic inconsistencies with respect to the final performance of the DIVA satellite. 
The effect of such cosmic scatter probably is that the final performance of DIVA might be less accurate than our results from these ideal synthetic spectra, or, that the accuracy curves of Fig.\,\ref{Fno2_ext} and Fig.\,\ref{G_ext} are to be shifted somewhat to brighter magnitudes.

We argued that additional UV data can improve the parameter results considerably in the classification process. The spectral library of our present simulations does not include, however, changes in, e.g., alpha-process elements which can show up in changes also in the range of DIVA's UV channels (e.g. the CN violet system in the range from 385 to 422 nm). 

The conclusions of the discussion are\\
1) The ANN method is well suited to obtain astrophysical parameters 
from DIVA \dispis.\\
2) The accuracy obtained is related with the strength of the signal of each 
parameter as present in a \dispi: 
\teff\ is best, followed by \logg, and then \ebv\ and \feh.\\
3) The accuracy is clearly related with temperature: toward higher 
temperature the signal of both \logg\ and \feh\ decreases considerably. \\
4) Our results were obtained with synthetic spectra. Real stars will not 
all behave like text book objects and the classification coming from real data 
will necessarily be less good, albeit always to an unknown amount per star.\\
5) The classification quality is absolutely adequate to be able to select 
objects of desired characteristics from the final DIVA database 
to do statistical analyses and/or 
for efficient post mission type-related investigations.

\begin{acknowledgements}
This project is carried out in preparation for the DIVA mission and we thank
the DLR for financial support (Projectnr. 50QD0103).  We thank Martin Altmann, Uli Bastian, Michael Hilker, Thibault Lejeune, Valeri Malyuto and Klaus Reif 
for helpful discussions. We also thank Oliver Cordes, Ole
Marggraf and Sven Helmer for assistance with the computer system level support.
\end{acknowledgements}

\bibliographystyle{apj}
\bibliography{H4080.bib}

\end{document}